\documentclass[useAMS,usenatbib]{mn2e}
\usepackage{epsfig}
\usepackage{graphicx}
\usepackage{journal_names}
\pdfoutput=1
\usepackage[english]{babel}
\usepackage{fancyhdr,txfonts} 
\usepackage{hyperref}
\usepackage{natbib}
\usepackage{graphicx}
\usepackage{sidecap}
\usepackage{subfigure}
\usepackage{amssymb}
\usepackage{wasysym}

\def\gsim{\;\rlap{\lower 2.5pt\hbox{$\sim$}}\raise 1.5pt\hbox{$>$}\;}
\def\lsim{\;\rlap{\lower 2.5pt\hbox{$\sim$}}\raise 1.5pt\hbox{$<$}\;}

\usepackage[usenames,dvipsnames]{color}

\newcommand{\Rmnum}[1]{\expandafter\@slowromancap\romannumeral #1@}

\title[The Lyman-$\beta$ and Lyman-$\alpha$ curvature ratio]{Constraining the temperature--density relation of the intergalactic medium with the Lyman-\boldmath{$\alpha$} and \boldmath{$\beta$} forests }
\author[Boera et al.]{\noindent
Elisa Boera$^{1}$\thanks{E-mail: eboera@swin.edu.au}, Michael T.~Murphy$^{1}$, George D.~Becker$^{2,4}$, James S.~Bolton$^{3}$
\\~\\
$^1$ Centre for Astrophysics and Supercomputing, Swinburne University of Technology, Hawthorn, Victoria 3122, Australia\\
$^2$ Space Telescope Science Institute, 3700 San Martin Dr, Baltimore, MD 21218, USA\\
$^3$ School of Physics and Astronomy, University of Nottingham, University Park, Nottingham, NG7 2RD\\
$^4$ Department of Physics \& Astronomy, University of California, Riverside, 900 University Avenue, Riverside, CA,92521, USA\\
}

\begin{document}

\date{Accepted 2015 October 28.  Received 2015 October 27; in original form 2015 June 20. }

\pagerange{\pageref{firstpage}--\pageref{lastpage}} \pubyear{2014}

\maketitle

\label{firstpage}

\begin{abstract}
The post-reionization thermal state of the intergalactic medium
is characterized by a power-law relationship between temperature and
density, with a slope determined by the parameter $\gamma$.  We
describe a new method to measure $\gamma$ using the ratio of flux
curvature in the Lyman-$\alpha$ and $\beta$ forests.  At a given
redshift, this curvature ratio incorporates information from the
different gas densities traced by Lyman-$\alpha$ and $\beta$
absorption. It is
relatively simple and fast to compute and appears robust against
several observational uncertainties.  We apply this
technique to a sample of 27 high-resolution quasar spectra from the
Very Large Telescope. While promising statistical errors on $\gamma$ appear
to be achievable with these spectra, to reach its full potential, the
dependence of the curvature ratio on the thermal state of the gas in
the foreground Lyman-$\alpha$ forest will require further, detailed
forward modelling.

\end{abstract}

\begin{keywords}
intergalactic medium -- quasar: absorption lines -- cosmology: observations 
\end{keywords}

\section{Introduction}\label {sec:intro}

The intergalactic medium (IGM) at low densities, with overdensities
$\Delta=\rho / \bar{\rho}\lesssim 10 $ at $z \sim 2$-$4$, where
$\bar{\rho}$ is the gas mean density, has a long adiabatic cooling
time that allows it to maintain a record of important events that
affected its thermal history, such as H{\sc \,i} and He{\sc \,ii}
reionisation \citep{HuiGnedin1997}.  The energy injected on relatively
short timescales during these epochs will increase the IGM
temperature, but it should also change its temperature--density
($T$--$\rho$) relation (e.g.~\citealt{Ricotti00}; \citealt{Schaye00}). In the
simplest scenario, the interplay between cooling and photoionization
heating by the UV background (UVB) results in a well-defined
$T$--$\rho$ relation, $T(\Delta)=T_{0}\Delta^{\gamma -1}$, with a
power-law slope, $\gamma-1$, and a temperature $T_0$ at the cosmic
mean density (\citealt{HuiGnedin1997,McQuinn2015}).
 
However, during and immediately after the H{\sc \,i} and He{\sc \,ii}
reionisations (at $z=6$--11 and 3--4, respectively, e.g.~\citealt{McGreer15}; \citealt{Syphers14}; \citealt{Worseck14}) cosmological simulations
predict that $\gamma$ may vary, becoming multi-valued and
spatially-dependent (e.g.~\citealt{Bolton04}; \citealt{McQuinn09}; \citealt{Meiksin12}; \citealt{Compostella13}; \citealt{Puchwein15}). Despite
considerable recent improvements, accurately simulating the effect of
reionisation events on the IGM remains an open challenge. Precise observational constraints are therefore desirable.

The main laboratory to detect variations in the $T$--$\rho$ relation
has been the H{\sc \,i} Lyman-$\alpha$ (Ly$\alpha$) forest in quasar spectra.
Efforts to infer the thermal state of the IGM have used either
line-profile decomposition to measure gas temperature as a
  function of column density
(e.g.~\citealt{Schaye00}; \citealt{Ricotti00}; \citealt{McDonald01}; \citealt{Rudie13}; \citealt{Bolton13}) or a
variety of statistical approaches which are valuable at higher
redshifts, $z>3$, where line fitting is problematic
(e.g.~\citealt{Theuns02}; \citealt{Becker07}; \citealt{Bolton08}; \citealt{Lidz10}; \citealt{Becker11}; \citealt{Boera14}; \citealt{Lee15}).
While these methods probe wide redshift and density ranges
($z\approx1.6$--5, $\Delta\approx0.3$--8), large
uncertainties, particularly in the measurements of $\gamma$, remain.

One way to improve this situation is to constrain the
$T$--$\rho$ relation by comparing Ly$\alpha$ and higher-order
Lyman-series transitions, such as Lyman-$\beta$ (Ly$\beta$). Ly$\beta$ lines of
moderate optical depth ($\tau\sim 0.1$--1.0) arise from higher
overdensities where Ly$\alpha$ lines may be saturated; the
Ly$\alpha$-to-$\beta$ optical depth ratio is
$f_{\alpha}\lambda_{\alpha}/f_{\beta}\lambda_{\beta}=6.24$
(proportional to the ratios of oscillator strengths and rest
wavelengths). 
Indeed, using the Ly$\beta$ forest in IGM temperature measurements
has been suggested in several theoretical works (e.g.~\citealt{Dijkstra04}; \citealt{Furlanetto09}; \citealt{Irsic14}).  However, so far no
practical attempt has been made to directly measure $\gamma$ from a
joint Ly$\alpha$ and $\beta$ forest analysis. One challenge is that
the Ly$\beta$ forest is entangled with lower-redshift, foreground Ly$\alpha$ absorption. Hereafter, we refer to the region between the Ly$\beta$ and Ly$\gamma$ emission lines, where the Ly$\beta$ and foreground Ly$\alpha$ absorption occurs, as the
Ly$\beta$$ +$$\alpha$ region. However, assuming that the
Ly$\beta$ and foreground Ly$\alpha$ lines arise from physically uncorrelated IGM
structures, a possible strategy to overcome this problem is to
statistically compare the properties of the Ly$\beta$$+$$\alpha$ and
Ly$\alpha$ regions.

In this Letter we present a new method to constrain the slope of the
$T$--$\rho$ relation using the two forest regions (Ly$\alpha$ and
Ly$\beta$$ +$$\alpha$) in 27 high resolution quasar spectra.  We use
a statistic based on the flux curvature analysis of \citet{Becker11}
and \citet{Boera14}. These previous works demonstrated that the
curvature method can measure the temperatures at the (redshift
  dependent) characteristic densities probed by the Ly$\alpha$
forest. However, as only a narrow density range is constrained, it has
not yet been used to measure the slope of the $T$--$\rho$ relation
from data \citep[but see][for a recent theoretical analysis using the
 Ly$\alpha$ curvature distribution]{Padmanabhan15}. Using
hydrodynamical simulations, we show that the ratio between the
curvatures of corresponding Ly$\alpha$ and Ly$\beta$$+$$\alpha$
forest regions (i.e.~around the same redshift, where in the Ly$\beta$$+$$\alpha$ region the redshift refers to Ly$\beta$ only) is sensitive to differences in the IGM thermal state
between the two density regimes. Averaged over many lines of sight,
this curvature ratio allows $\gamma$ to be measured with little
sensitivity to $T_{0}$. 

We demonstrate the potential for this technique using 27 quasar
spectra spanning the redshift range $2\lesssim z\lesssim 3.8$. That
potential is currently limited by assumptions regarding the thermal state of the foreground Ly$\alpha$ forest; we propose how these can be
overcome with future simulations and a refined data analysis approach.

\section{The observational data}\label {sec:data}

The 27 quasar spectra were originally retrieved from the archive of
the Ultraviolet and Visual Echelle Spectrograph (UVES) on the Very
Large Telescope (VLT). They were selected on the basis of quasar
redshift, wavelength coverage and signal-to-noise ratio from the
sample of 60 spectra used in \citet{Boera14} (hereafter B14). 
They have resolving power
$R\sim50000$ and continuum-to-noise $\ge$24\,pix$^{-1}$ in the
Ly$\alpha$ forest region. This
level of spectral quality is necessary so that the curvature
measurement is not dominated by noise and unidentified metal
lines. Because our new method compares the curvature of the Ly$\alpha$ and
$\beta$$+$$\alpha$ forest regions, we extended these same criteria to
the Ly$\beta$$+$$\alpha$ region at $z\sim2.0$--3.8, reducing the sample
from 60 to 27 spectra. The quasar sample details are provided in Table
1 in the Supporting Information (hereafter SI). 

To establish an initial continuum in the Ly$\beta$$+$$\alpha$ region,
we applied the same automatic, piece-wise polynomial fitting algorithm in B14, with the same parameters, leaving the Ly$\alpha$ region unchanged. While manual refitting was necessary in some spectra for
particular parts of the Ly$\beta$$+$$\alpha$ region, this initial
continuum has little effect on the final curvature measurements
because the spectra are subsequently renormalized based on a b-spline fit to the flux profile (see Section~\ref{subsec:simA}).

\section{The simulations}\label{sec:sim}

We used the same hydrodynamical simulations in B14 to
produce synthetic spectra for Ly$\beta$$+$$\alpha$ and Ly$\alpha$ in
the redshift range $z=2.0$--3.5. The {\sc gadget-3} smoothed-particle
hydrodynamic simulations include dark matter and gas, with $2 \times
512^{3}$ particles and a gas particle mass of $9.2\times 10^{4}
M_{\odot}$ in a periodic box of 10 comoving $h^{-1}$ Mpc
(see B14 for details). The gas, assumed to be optically
thin, is in equilibrium with a spatially uniform UVB \citep{Haardt01},
but the photoheating rates have been rescaled so that the
corresponding values of $T_{0}$ and $\gamma$ vary between different
simulations. The parameters characterizing the different models are
summarized in Table 2 in the SI. From each model,
synthetic Ly$\alpha$ forest spectra were generated for 1024 random
lines of sight through each of 6 redshift snapshots over the range
$z=2.0$--3.5. The Ly$\beta$$+$$\alpha$ spectra were produced by
scaling the Ly$\alpha$ optical depths by a factor
$f_{\alpha}\lambda_{\alpha}/f_{\beta}\lambda_{\beta}=6.24$ and
contaminating them with randomly selected foreground Ly$\alpha$
absorption from lower redshift outputs ($z=1.6$--2.7) from the
  same simulation.

Finally, as in B14, we calibrated the synthetic spectra to
match the properties of the real spectra (i.e.~resolving power, pixel
size and signal-to-noise ratio). The only difference is that, in this
work, we scaled the effective optical depth of the synthetic spectra
to match the recent results from \cite{Becker12} rather than to a
direct measurement from our spectra. As shown in
Section~\ref{subsec:simA}, our measurements are relatively insensitive
to this optical depth calibration.

Box size and mass resolution convergence tests are available in Fig.~1
and Table 3 of the SI.

\begin{figure} 
\begin{center}
\includegraphics[width=0.96\columnwidth]{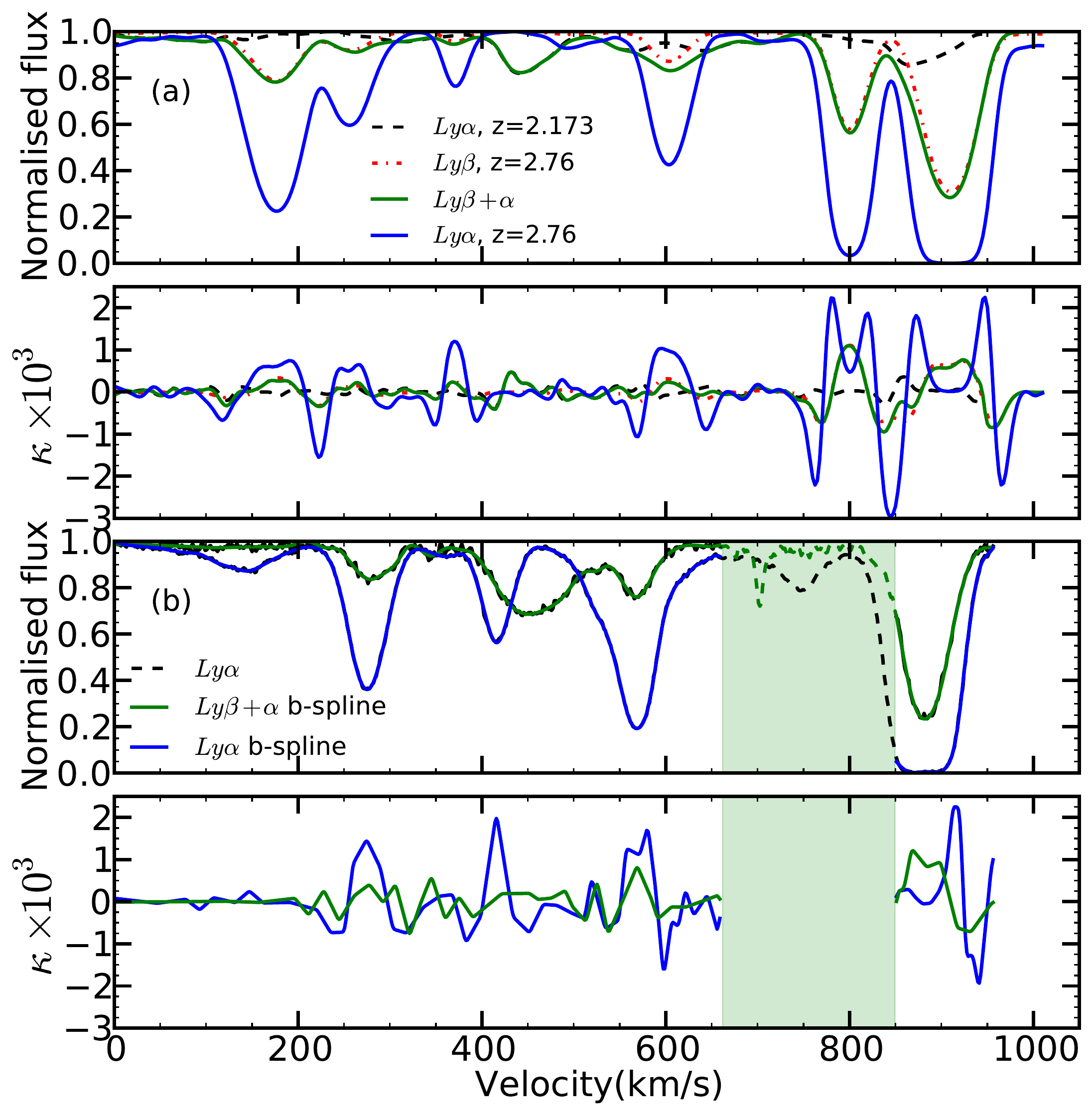} 
\vspace{-0.8em}
\caption{\small Curvature for simulated and real spectra. The top
  two panels display the simulated Ly$\alpha$ (blue solid line) and
  Ly$\beta$$+$$\alpha$ forests (green solid line) at $z=2.76$ and the
  curvature from b-spline fits.  The spectra are obtained by
  contaminating the corresponding Ly$\beta$ forest (red dotted line)
  with a randomly chosen Ly$\alpha$ section at lower redshift (black
  dashed line). The lower two panels are as above but for a real Ly$\alpha$ and Ly$\beta$$+$$\alpha$ spectrum. The spectra (black
  lines) are plotted behind the b-spline fits. Shading shows a
  Ly$\beta$$+$$\alpha$ region contaminated by metal absorption (green
  dashed line); the corresponding part of the Ly$\alpha$ spectrum is
  also masked (black dashed line).}
\label{fig:Ksim}
\end{center}
\end{figure}

\section{The curvature ratio method}\label {sec:method}

The curvature for a region of spectrum is defined as $\kappa\equiv
F^{\prime\prime}(1+{F^\prime}^2)^{-3/2}$ \citep{Becker11}, with the
first and second derivatives of the normalised flux (i.e.~transmission; $F^\prime$,
$F^{\prime\prime}$) taken with respect to wavelength or relative
velocity. As demonstrated in previous works, the Ly$\alpha$ forest
curvature is directly related to the IGM temperature at the
characteristic overdensities probed by this absorption ($\Delta_{\alpha}\simeq5$--2 for $z=2$--3.5), regardless of
$\gamma$. Because the median overdensity contributing to Ly$\beta$
forest absorption is higher than that for Ly$\alpha$
\citep{Furlanetto09}, the mean absolute curvature computed from sections of Ly$\beta$$+$$\alpha$ forest will be, on average, a tracer of the IGM temperature in a higher density regime ($\Delta_{\beta}\simeq8$--5 for $z=2$--3.5). Therefore, the curvature ratio, at each redshift $z$,
\begin{equation} \label{R}
R_{\kappa}(z)\equiv\frac{\langle|\kappa_{\beta+\alpha}(z)|\rangle}{\langle|\kappa_{\alpha}(z)|\rangle}\,,
\label{eq:ratio}
\end{equation}
probes temperatures at two different gas densities and,
consequently, is sensitive to $\gamma$. Here, the mean absolute curvatures for
Ly$\beta$$+$$\alpha$ and Ly$\alpha$ are averaged over spectral sections of 10 comoving $h^{-1}$\,Mpc centered on the same redshift (corresponding to the
simulation box size and, again, $z$ in the Ly$\beta$$+$$\alpha$ region refers to Ly$\beta$ absorption). Due to the presence of  foreground Ly$\alpha$ absorption, $R_{\kappa}(z)$ also depends on the evolution of $\gamma$ and $T_{0}$ with $z$ (see Section \ref{subsec:SystErr}).

\subsection{Analysis of simulated spectra: the $\gamma$--$\log\langle R_{\kappa}\rangle$ relation} \label {subsec:simA}

We find the connection between the curvature ratio, $R_{\kappa}$, and
$\gamma$ at the redshifts of the simulation snapshots ($z$=[2.173,
  2.355, 2.553, 2.760, 3.211, 3.457]) by computing the mean
$R_{\kappa}$ over the 1024 simulated lines of sight for each thermal
history, and fitting a simple function between $\log\langle
R_{\kappa}\rangle$ and $\gamma$.

Single $R_{\kappa}$ values for each line of sight and redshift were
obtained as defined in Eq.~(\ref{eq:ratio}). To compute the mean
absolute curvature of synthetic Ly$\alpha$ and Ly$\beta$$+$$\alpha$
sections, we adopt exactly the same algorithm and parameters described in \citet{Becker11}
and B14, an example of which is shown in
Fig.~\ref{fig:Ksim}: in each section of artificial spectrum, the mean
absolute curvature is computed from a cubic b-spline fit which has been
re-normalized by its maximum value in that interval. This approach is required to avoid
systematic errors when determining the curvature from real spectra, so
it must be applied to the synthetic spectra for consistency. The
b-spline fit reduces the sensitivity of the curvature to noise and the
re-normalization minimizes potential uncertainties arising from
inconsistent continuum placement. Finally, only the pixels where the
re-normalized b-spline fit falls in the range 0.1--0.9 are used to
measure the mean absolute curvature of each section. In this way we maximize the sensitivity to the signal and avoid possible systematic uncertainties:
we exclude saturated pixels, which do not contain useful information, and
pixels with little-to-no absorption whose curvature is near zero and
uncertain.

\begin{figure} 
\begin{center}
\includegraphics[width=0.90\columnwidth]{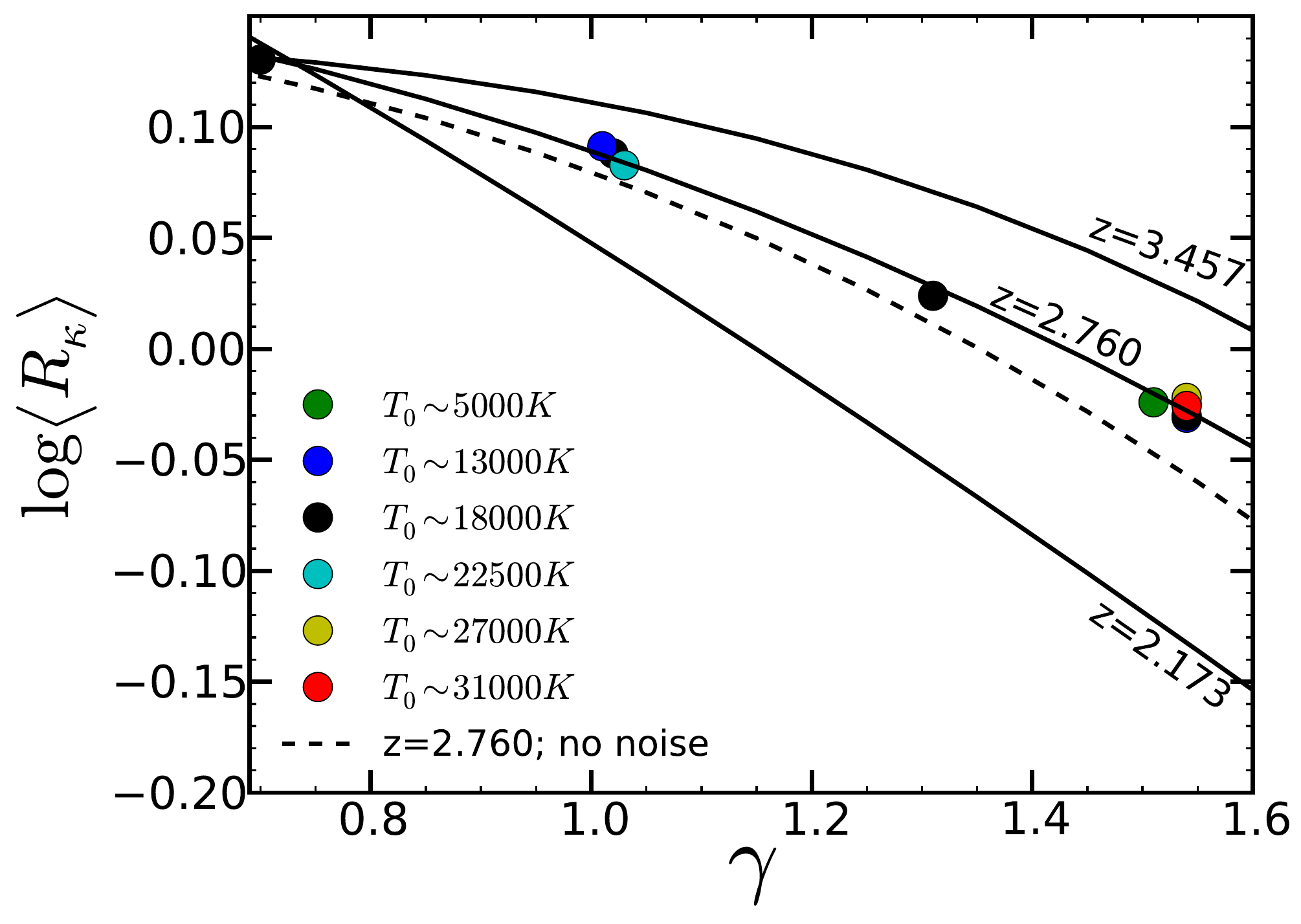} 
\vspace{-1.2em}
\caption{\small Relationship between $\gamma$ and the mean curvature ratio,
  $\log\langle R_{\kappa}\rangle$, from our nominal simulations at 3 different redshifts. For clarity, we show the values of $\log\langle R_{\kappa}\rangle$ from the different models and the relation without noise for $z=2.760$ only (coloured points and dashed line, respectively). See Fig.~2 and Table 4 in the SI for the plots corresponding to the other redshifts and the fitting parameters.}
\label{fig:Rg}
\end{center}
\end{figure}

In Fig.~\ref{fig:Rg} we present the $\gamma$--$\log\langle
R_{\kappa}\rangle$ relationship obtained by averaging the curvature
ratio computed from the synthetic spectra with different thermal
histories, i.e. different $\gamma$ and $T_{0}$ values.  
At each redshift, models characterized by the same $\gamma$ value, but with $T_{0}$ ranging between 5000--31000\,K, yield $\log\langle R_{\kappa}\rangle$ values that vary by only $\lesssim0.01$\,dex. Therefore, for each redshift, it is possible to fit a simple power-law ($a\gamma^{b}+c$) that connects the mean $\log\langle R_{\kappa}\rangle$ and
$\gamma$ independently of $T_{0}$, at least in these simulations (see also Table 4 and Fig.~2 in the SI). As redshift decreases, the relation
between $\log\langle R_{\kappa}\rangle$ and $\gamma$ becomes steeper, as shown in Fig.~\ref{fig:Rg}. Given this tight correspondence, the curvature ratio represents an interesting tool to 
independently measure the
slope of the IGM $T$--$\rho$ relation.

The sensitivity of this nominal $\gamma$--$\log\langle
R_{\kappa}\rangle$ relation to observational uncertainties in the
spectra was tested as follows:
\begin{itemize}
\item{Noise}: The synthetic spectra from which we obtain the nominal
  relationship need to include noise at the same level as in the real
  spectra. If, as an extreme example, no noise was added,
  Fig.~\ref{fig:Rg} shows the effect on the $\gamma$--$\log\langle
  R_{\kappa}\rangle$ relation: a $\sim$5\,per cent decrease in $\langle R_{\kappa}\rangle$ ($\sim0.02$ dex in $\log\langle R_{\kappa}\rangle$). This would cause a
  $\sim$8\,per cent underestimation of $\gamma$, comparable to the
  statistical errors in our observational sample (see Section \ref{subsec:dataA}). Therefore, any errors in how the
  noise properties of a sample like ours are incorporated into the $\gamma$--$\log\langle
  R_{\kappa}\rangle$ relation should cause relatively small systematic uncertainties in $\gamma$ measurements.
\item{Effective optical depth}: Changing the effective optical depth
  by 10\,per cent alters $\log\langle R_{\kappa}\rangle$ by 0.01 dex and, consequently, $\gamma$ by $\sim$4\, per cent,
  again well within the statistical uncertainties in our observational sample. This point is
  particularly promising because the curvature for the Ly$\alpha$
  forest alone (and the Ly$\beta$$+$$\alpha$ forest alone) is
  considerably more sensitive to this quantity, as explored in B14.
\end{itemize}
However, while the nominal $\gamma$--$\log\langle R_{\kappa}\rangle$
relation is robust to these observational aspects,
systematic uncertainties in the evolution of the IGM thermal state must also be considered (see Section
\ref{subsec:SystErr}).

\subsection{The observed curvature ratio}\label {subsec:dataA}

To apply the method to the 27 real quasar spectra we compute the
curvature ratio in sections of 10 comoving $h^{-1}$Mpc of metal-free
Ly$\alpha$ and corresponding Ly$\beta$$+$$\alpha$ forest regions.
Narrow metal lines ($b\lesssim$15\,km$s^{-1}$) represent a potentially
serious source of systematic errors in any measure of forest
absorption and should be avoided. With this aim we ``clean" the
spectra, using the exact procedure detailed in B14, extended to the Ly$\beta$$+$$\alpha$ region: metal absorbers
redward of the Lyman-$\alpha$ emission line are identified and all
strong metal transitions at their redshifts are masked out, followed
by a by-eye check of the remaining forest.

While the metal correction produces spectra that are reasonably free
of contaminants, this procedure reduces the quantity of information
available in different sections in a non-uniform way, introducing a
possible source of bias.  Because the curvature ratio traces
differences in the absorption features of two different regions of the
same observed spectrum, avoiding systematic effects requires that they
cover the same absorption redshift range. Therefore, before measuring
$R_{\kappa}$ from the real data, we mask out regions of the
Ly$\beta$$+$$\alpha$ forest corresponding to any range masked from
the Ly$\alpha$ forest, and vice versa. Figure \ref{fig:Ksim} shows an
example of this masking procedure. Finally, possible edge effects are
avoided: we do not include the 4 pixels closest to the edge of any
masked region in the curvature ratio calculation.

\begin{figure} 
\begin{center}
\includegraphics[width=0.98\columnwidth]{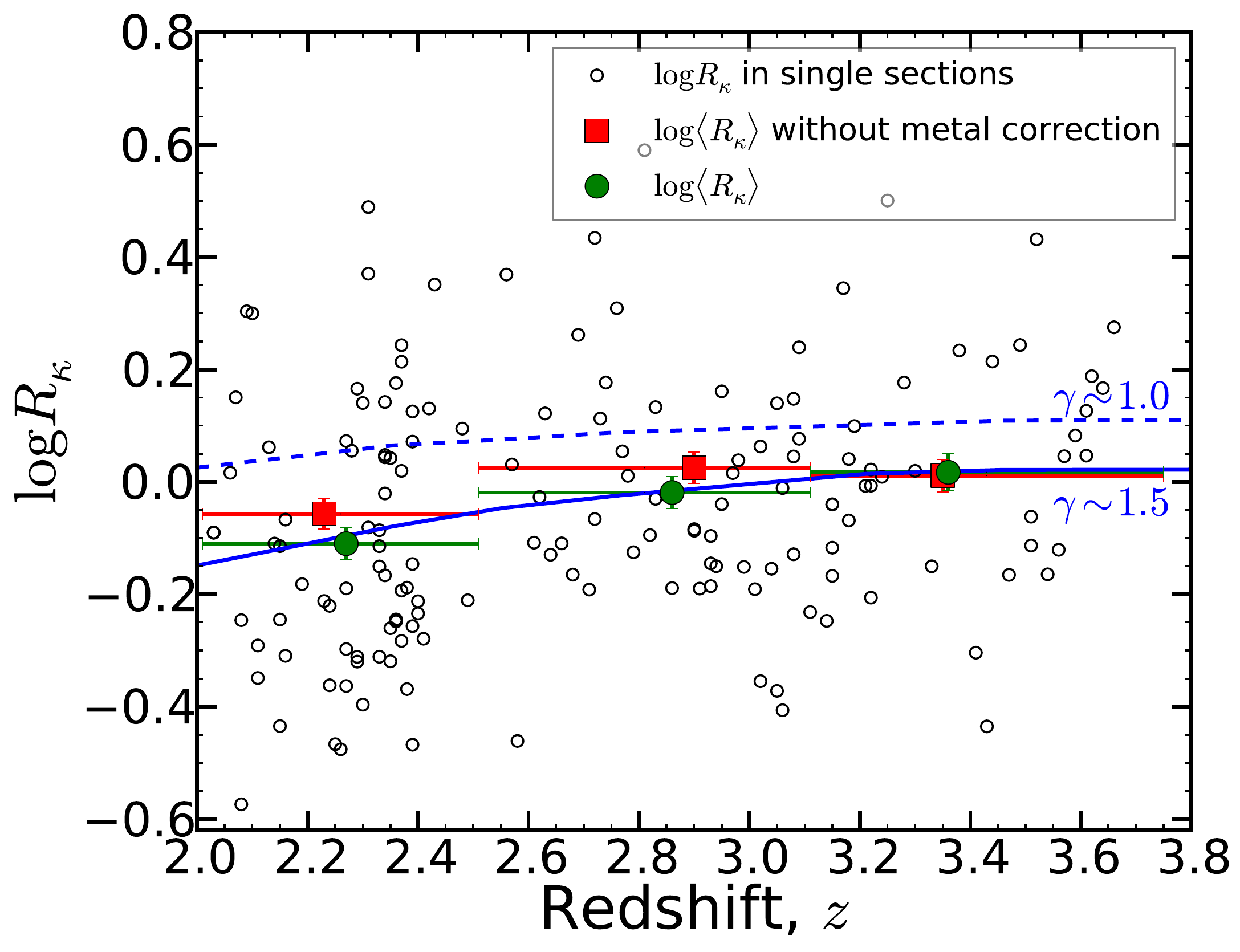} 
\vspace{-0.8em}
\caption{\small Curvature ratio, $R_{\kappa}$ from each pair of
  Ly$\beta$$+$$\alpha$ and Ly$\alpha$ sections from 27 UVES/VLT
  quasar spectra (black circles). The measurement of $\log\langle
  R_{\kappa} \rangle$, averaged within each redshift bin, is shown
  with 1-$\sigma$ bootstrap errors both before (red squares) and after
  (green points) the metal masking procedure. The expected evolution
  of $\log \langle R_{\kappa} \rangle$, from our nominal simulations
  with relatively constant $\gamma$, is presented for $\gamma\sim1.5$
  (solid blue line) and $\gamma\sim1.0$ (dashed blue line). The
  redshift bins are $(z_{\rm min},\bar{z},z_{max}) = (2.0,2.27,2.5),
  (2.5,2.86,3.1), (3.1,3.36,3.74)$, and the (metal-masked) $\log
  \langle R_{\kappa} \rangle$ measurements are $-0.110\pm0.026$,
  $-0.019\pm0.030$ and $0.017\pm0.032$, respectively.}
\label{fig:Ratio}
\end{center}
\end{figure}

Figure~\ref{fig:Ratio} presents the curvature ratio results from our
observational sample. The statistical uncertainty in an individual $\log R_{\kappa}$ measurement,
computed from a single pair of Ly$\beta$$+$$\alpha$ and
Ly$\alpha$ spectral sections, is negligible compared to the much larger
variance among different measurements. Therefore, in each of three broad redshift
bins, which include $>$30 individual measurements, we calculate the
mean $\log \langle R_{\kappa} \rangle$ and its uncertainty using a bootstrap technique using the measurements in the bin. 
The width and roughly Gaussian shape of the $\log R_{\kappa}$ distribution
within each bin is also reproduced by our simulated spectra,
providing some confidence that the simulations adequately describe the
statistical properties of the observed forest absorption (see Fig.~3 in the SI).

There is evidence for a mild evolution in $\log \langle R_{\kappa} \rangle$
as a function of redshift. A Spearman rank correlation test reveals a
positive correlation ($r\approx0.26$) with an associated probability
of $\approx$0.001. For a constant $\gamma$, this increase in
$\log \langle R_{\kappa} \rangle$ is consistent with the expected increase
in $\log \langle R_{\kappa} \rangle$ at
increasing redshifts seen in Fig.~\ref{fig:Rg}. In particular, our
$\log \langle R_{\kappa} \rangle$ measurements in Fig.~\ref{fig:Ratio} show good agreement
with the expected evolution for a model with $\gamma \sim 1.5$, i.e.~a scenario consistent with the recent result of $\gamma=1.54\pm0.11$ at $z\sim 2.4$ from \cite{Bolton13}. The
absence of a strong change in the curvature ratio in the redshift
range considered is consistent with the assumption in the nominal
simulations that $\gamma$ remains relatively constant over redshifts
$z=2$--3.5 (but see discussion on systematic uncertainties below).

Figure~\ref{fig:Ratio} also shows $\log \langle R_{\kappa} \rangle$
computed without first masking the metal absorption lines. Even though
the effect of metal contamination is important when measuring the
curvature of the Ly$\alpha$ forest alone (B14), it is
similar in the corresponding sections of Ly$\beta$$+$$\alpha$ forest,
so the curvature ratio is less sensitive to this correction. The
results in Fig.~\ref{fig:Ratio} show that, even \emph{without}
applying the metal correction, the bias introduced in $\log \langle R_{\kappa} \rangle$ is $\sim 0.05$ dex. 
Therefore, possible errors in the metal masking procedure will introduce small uncertainties
in $\gamma$ compared to our statistical ones.
\subsection{Modelling uncertainties}\label{subsec:SystErr}

Our nominal simulations assume only mild evolution in $T_{0}$ ($\Delta
T_{0}\sim 2000\rm\,K$) and $\gamma$ ($\Delta \gamma \lesssim 0.02$) between the Ly$\beta$ and the foreground Ly$\alpha$ redshifts (see Fig. 1 in
\citealt{Becker11}). However, if these parameters vary more drastically on short timescales
due to e.g.~blazar heating (\citealt{Puchwein12}) or non-equilbrium photo-ionisation effects during He{\sc \,ii} reionisation \citep{Puchwein15}, these assumptions may have important
consequences for the $\gamma$--$\log\langle R_{\kappa}\rangle$
relation and, therefore, preclude a final conversion of our
$\log\langle R_{\kappa}\rangle$ measurements to $\gamma$ values 
with formal error bars.

\begin{figure}
\begin{center}
\includegraphics[width=0.98\columnwidth]{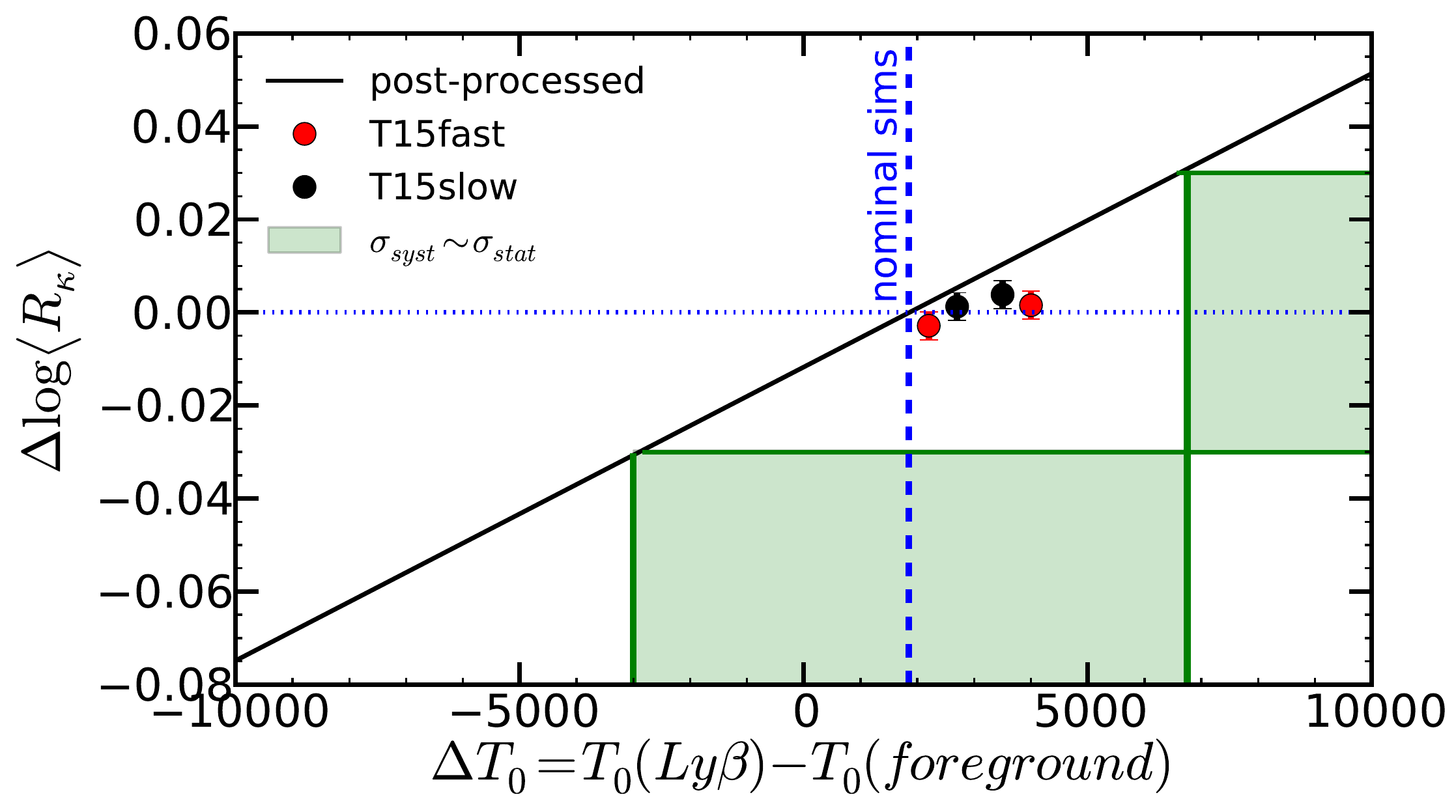} 
\vspace{-1.0em}
\caption{\small Expected relation between the variation in $\log\langle
  R_{\kappa}\rangle$ and the temperature change, $\Delta T_{0}$
  between the Ly$\beta$ redshift and foreground Ly$\alpha$ forest
  (black curve, see text for details). The green shading
  indicates the typical statistical uncertainty in our measured
  $\log\langle R_{\kappa}\rangle$ values. The $\Delta T_{0}$ in our
  nominal simulations is typically $\la$2000\,K (dashed line) and the
  specific results for the self-consistent `T15fast' and `T15slow'
  simulations with their statistical errors are shown for $z=2.7$ and
  3.2.}
\label{fig:T0var}
\end{center}
\end{figure}

We can construct a toy model to investigate how sensitive the
$\gamma$--$\log\langle R_{\kappa}\rangle$ relation is to rapid evolution
in $T_{0}$ and $\gamma$ by altering these parameters in the foreground Ly$\alpha$ forest
 via a simple post-processing of the simulated spectra. For a
given simulation model, at a given redshift, new synthetic spectra are
extracted after imposing a new one-to-one power-law $T$--$\rho$
relationship. Note this is an approximation, as it removes the
natural dispersion in the temperatures at a given overdensity
from shock-heating/radiative cooling. However, in this way we
can easily modify the $T_{0}$ and $\gamma$ parameters, and their
evolution, to explore the effect on the $\gamma$--$\log\langle R_{\kappa}\rangle$ without running new hydrodynamical simulations.
The largest effects on the $\gamma$--$\log\langle R_{\kappa}\rangle$
relation were found in the following two tests:
\begin{itemize}
\item Rapid evolution in $T_{0}$: For the redshift range $z=2$--3.5, $T_{0}$
  evolves in our nominal simulations such that the temperature
  difference between the foreground and the Ly$\beta$ redshift is
  small, $\Delta T_{0}\equiv T_{0}({\rm Ly}\beta)-T_{0}({\rm
    foreground}) \la 2000$\,K. To test the effect of much stronger
  variations in the temperature at the mean density, we modified the
  values of $T_{0}$ in each of our nominal simulations, for the
  foreground Ly$\alpha$ forest only, using the post-processing approach. We then computed the change in $\log\langle R_{\kappa}\rangle$ compared
  to the nominal values, $\Delta\log\langle
  R_{\kappa}\rangle$. Fig.~\ref{fig:T0var} shows the direct relationship between $\Delta T_{0}$ and $\Delta\log\langle
  R_{\kappa}\rangle$. For example, if $T_{0}$ changes by a
    further $\approx$5000\,K between $z=3.2$ (Ly$\beta$) and
  $z=2.6$ (foreground), we would expect a systematic error in our
  measurement of $\log\langle R_{\kappa}\rangle$ of $\approx$0.03,
  similar to the statistical error per redshift bin derived from our
  27 quasar spectra.
\item Rapid evolution in $\gamma$: Similar to the $\Delta T_{0}$ test above,
  we emulated rapid changes in $\gamma$ over short timescales, $\Delta z\approx0.6$, by post-processing the
  foreground Ly$\alpha$ forest only. Again, we find a strong correlation between the change in foreground $\gamma$ and
  $\log\langle R_{\kappa}\rangle$; for example, a decrease in the
  foreground $\gamma$ by 0.15 implies $\Delta\log\langle
  R_{\kappa}\rangle\approx0.03$, again equivalent to the statistical
  uncertainties in our 3 redshift bins.
\end{itemize}
On the other hand, we found that the $\gamma$--$\log\langle R_{\kappa}\rangle$ relation was
insensitive to variations in the integrated thermal history in the
simulations (i.e.~Jeans smoothing effects) when the instantaneous $T_0$ and $\gamma$ are
held fixed (see Fig.~4 in the SI). 

While highlighting the potential importance of assumptions for
the foreground Ly$\alpha$ forest, the above tests rely on a
rather simplistic toy model that does not reproduce self-consistently the evolution of the complex relationships between physical parameters.  Ideally, the sensitivity of the
$\gamma$--$\log\langle R_{\kappa}\rangle$ relation to different
physical assumptions and thermal histories would need to be tested
with additional self-consistent simulations in the redshift range of
interest.  Our simulation suite does offer one such self-consistent
test in the case of strong $T_{0}$ evolution: we used the `T15fast'
and `T15slow' simulations of \citet{Becker11} (see their Fig. 8) to
mimic a possible $\sim$5000\,K heating event from He{\sc \,ii}
reionisation at $z>3$ such that between the Ly$\beta$ and foreground
redshifts there was a typical decrease $\Delta T_{0}\approx
2000$--4000\,K. In both simulations we find that $\log\langle
R_{\kappa}\rangle$, varies by $\lesssim$0.01 compared to the nominal
simulations (see Fig.~\ref{fig:T0var}), which seems less sensitive to
substantial evolution in $T_{0}$ than implied by our simplistic toy
model.  However, differences in the pressure smoothing scale in these
models (which may be acting to improve the agreement) prevent us from
reliably estimating the systematic uncertainties involved without
further self-consistent tests. Nevertheless, these results
  suggest that the curvature ratio is a promising alternative tool to
  measure the density dependence of the IGM thermal state.

\section{CONCLUSIONS}

We have presented a new approach to constraining the slope of the
$T$--$\rho$ relation using the ratio of curvatures, $\langle
R_{\kappa}\rangle$, in the Ly$\alpha$ and $\beta$ forests. The statistic appears robust against observational uncertainties in the
  noise level, metal contamination and effective optical depth, and
is relatively simple and fast to compute. We measure $\langle
R_{\kappa}\rangle$ in 27 VLT/UVES quasar spectra, achieving $\approx$6\,per cent statistical accuracy in 3 redshift bins covering $z=2.0$--3.5. In the absence of any other systematics,
  this translates to a $\lesssim$10\,per cent uncertainty in $\gamma$, suggesting that $\langle
R_{\kappa}\rangle$ should be a useful tool to constrain the slope of the $T$--$\rho$ relation, possibly competitive with recent
  attempts to measure $\gamma$ using line decomposition
  \citep{Rudie13,Bolton13}.

The primary goal of this Letter is to introduce the $\langle
R_{\kappa}\rangle$ statistic.
However, we emphasize that the thermal state of the foreground Ly$\alpha$ forest may complicate a direct translation of $\langle
R_{\kappa}\rangle$ into a constraint on $\gamma$. The simulations used in this work have simple thermal histories where both $T_{0}$ and $\gamma$ remain relatively constant over $2\lesssim z\lesssim 5$. The impact on $\langle
R_{\kappa}\rangle$ for some deviations from these histories are explored in Section \ref{subsec:SystErr}; however, we have not attempted to capture the full range of possible $T_{0}$ and $\gamma$ evolutions expected for
non-equilibrium photo-heating during He II reionisation (e.g.~\citealt{McQuinn09}; \citealt{HM12}; \citealt{Puchwein15}) or more exotic
models incorporating e.g.~blazar heating (\citealt{Puchwein12}).  We
intend to more fully explore these issues in a future work. If the major systematic uncertainties can be marginalised over, or if $\langle
R_{\kappa}\rangle$ can be combined with other statistics to simultaneously determine
$T_{0}$(z) and $\gamma$(z), then the curvature ratio promises to be an  effective tool for constraining the thermal history of the IGM over a large
redshift range. 


\section*{Acknowledgements}
The simulations were performed using the Darwin Supercomputer of the
University of Cambridge High Performance Computing Service, provided
by Dell Inc.~using Strategic Research Infrastructure Funding from the
Higher Education Funding Council for England. MTM thanks the
Australian Research Council for \textit{Discovery Project} grant
DP130100568, GDB acknowledges support from the Kavli Foundation and
JSB acknowledges the support of a Royal Society University Research
Fellowship.

\small
\itemindent -0.48cm
\bibliography{ref}


\section*{Supporting Information}

Additional Supporting Information may be found in the online version of this article:\\
\textbf{SupportingInformation.pdf}

\label{lastpage}

\end{document}